A Computational Method for Studying Vibrational Mode Dynamics

Andrew Rohskopf [1], Ruiyang Li[2], Tengfei Luo[2], Asegun Henry[1]

[1] Massachusetts Institute of Technology, Department of Mechanical Engineering, Cambridge, MA, 02139, USA

[2] Department of Aerospace and Mechanical Engineering, University of Notre Dame, Notre Dame, IN 46556, United States

**Abstract**

The traditional picture of heat transfer in solids by atomic vibrations, also known as phonons, involves phonons scattering with each other like gas particles and is commonly referred to as the phonon gas model (PGM). This physical picture accounts for interactions among propagating (i.e., plane wave modulated) vibrational modes in an ideal crystal, but it becomes problematic when describing non-propagating modes arising in realistic non-idealized systems. Here, we introduce a more general formalism for studying phonon transport, which involves projection of the interatomic interactions themselves (i.e., not just the atom motion), onto the normal modes of the system. This shows, for the first time, how energy is exchanged between modes in real-time during molecular dynamics (MD) simulations, as opposed to other MD methods which use inferences based on correlations, or other time averaged schemes that do not preserve specific features in the real-time dynamics. Applying this formalism to the example case of modes interacting in a superlattice, we illustrate a new perspective on how phonon transport occurs, whereby individual normal modes share energy through specific channels of interaction with other modes. We also highlight that while a myriad of interaction pathways exist, only a tiny fraction of these pathways actually transfer significant amounts of energy, which is surprising. The approach allows for the prediction and simulation of these mode/phonon interactions, thus unveiling the real-time dynamics of phonon behavior and advancing our ability to understanding and engineer phonon transport.

**Introduction**

In solids and rigid molecules, atoms vibrate about their respective equilibrium positions, and this thermal motion can be understood as a superposition of the structure's normal modes, which are collective motions of atoms vibrating at individual frequencies. A wide variety of modes arise in different systems, with examples shown in Figure 1, that depend on molecular structure, the masses of atoms, and the stiffness of their chemical bonds/interactions. The procedure for calculating the mode frequencies and eigenvectors in a general large supercell of atoms is explained in the Methods section.

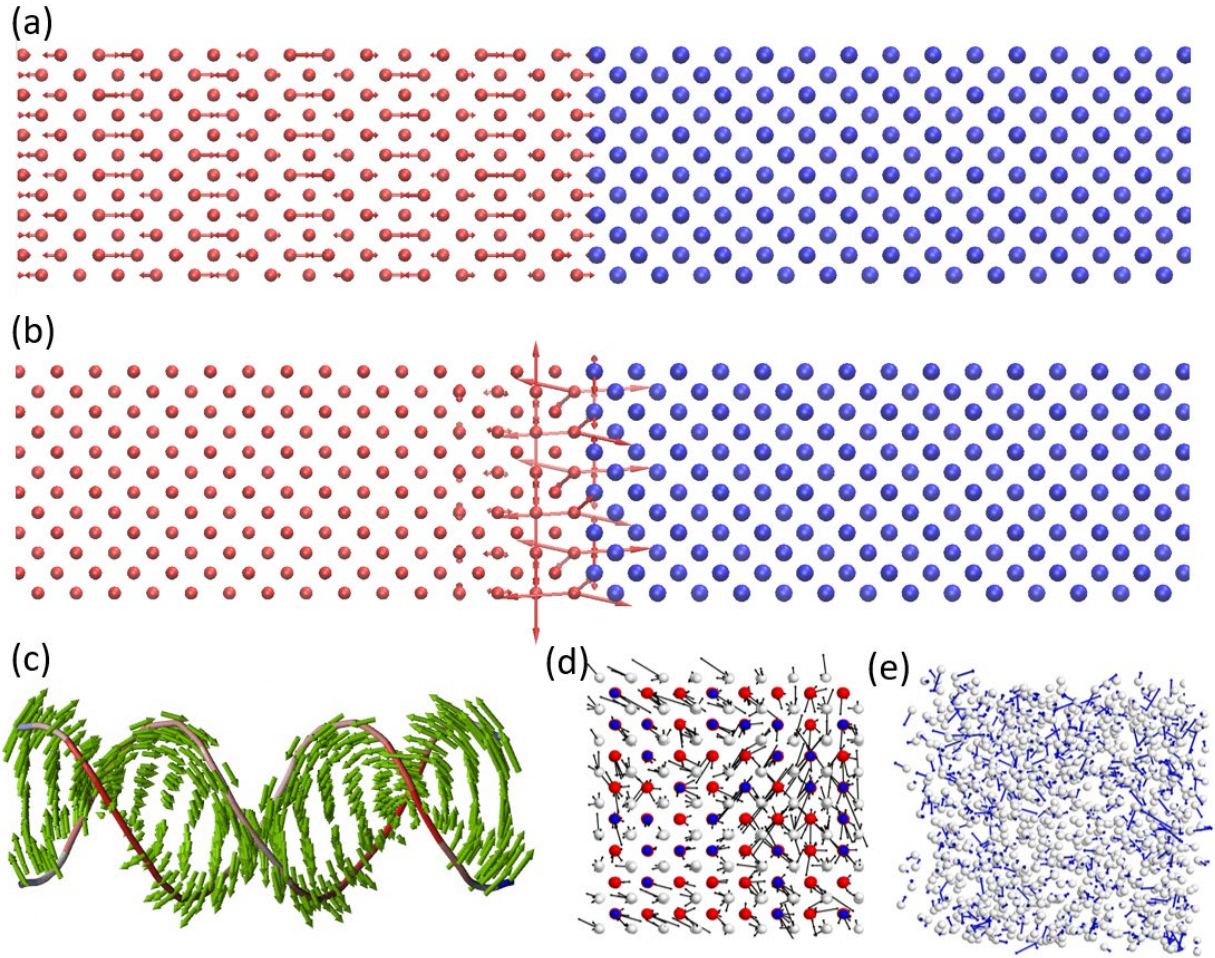

Figure 1 – Examples of normal modes for a variety of systems, illustrated by their eigenvectors (arrows) plotted at equilibrium atom positions. (a) A propagating mode on the silicon side of a silicon-germanium (Si-Ge) interface; such periodic modes arise in ideal crystalline materials and combine with other propagating modes to form wave-packets that transport energy/heat.[1,2] (b) A localized interface mode at the interface of a Si-Ge superlattice; the role of these modes in interfacial heat transfer has been inferred from correlation-based studies, but the mechanism of energy transfer remains unclear.[3] (c) Modes also arise in biological systems, such as this DNA molecule[4], and play roles in many biological mechanisms.[5-7] (d) and (e) show the diffusive-like modes arising in disordered alloys[8] and amorphous solids[9], respectively. While modes arise in a variety of systems and are readily studied within the framework of normal mode analysis[10] or lattice dynamics[11], interactions between modes remain unexplored aside from correlation-based methods which infer interactions[3,9].

Normal modes of vibration, which are hereby used interchangeably with the term phonons[12], exchange energy with each other and play pivotal roles in various electronic, chemical, and energy transfer phenomena by coupling with electrons or external fields to induce chemical reactions or phase transformations. It is in this sense that understanding phonon interactions has implications far beyond the field of phonon transport itself. This is because understanding the nature of phonon-phonon interactions lays a foundation for understanding how phonons couple to electronic and chemical phenomena. In theory, this could enable a new means of control over the properties of matter[13-15], such as catalyzing chemical reactions[16-18], phase transformations[19,20], or ion transport[21,22].

This naturally raises several questions, such as: how do modes transfer energy to other modes and why are certain mode-mode interactions are stronger than others? Do phonon interactions exhibit any special time resolved features, e.g., such as a special set of interactions that becomes activated when a phonon wave packet collides with an interface? Is there a useful descriptor that can provide insight on transport without requiring a dynamic simulation? These and many other questions can potentially be answered with a method for evaluating phonon dynamics, where the interactions between vibrational modes are simulated/predicted. Previous methods typically use molecular dynamics (MD) simulations to calculate the resulting mode amplitudes at each timestep and infer physical mechanisms of transport using mode-mode correlations[9]. However, it is important to note that correlation does not equate to causation, and therefore, even though methods such as Green Kubo Modal Analysis (GKMA)[9] and Interface Conductance Modal Analysis (ICMA)[3] along with others[23,24] measure correlations, interpreting these correlations as interactions is still conjecture/inference. Although it has been proffered as a proxy for mode-mode interactions, this has not actually been proven, and thus it is desirable to devise a method for direct evaluation of mode-mode interactions, which could confirm or deny such inference. We therefore introduce a formalism and method to directly calculate interactions and energy transfers among modes.

To understand this conceptually, consider the possibility that modes interact under some set of governing relations that determine the extent to which modes will influence each other's amplitudes. In essence, just as you have interatomic forces, there may be inter-mode forces that drive energy transfer between the equivalent of "nearest neighboring" modes, as opposed to nearest neighboring atoms. By this, what is meant is that atoms exert stronger forces on some atoms as opposed to others, and generally the interactions decay with increasing distance. One fundamental question is then: What is the mode level analog? Which modes exert the strongest influences/forces on each other?

To answer these questions, we consider bypassing the atom trajectory altogether and instead, seek to write the equations of motion entirely in terms of the mode-mode interactions. If this can be done, one could conceivably conduct the entire MD simulation in mode coordinates, and it would reveal instantaneous information about which modes are interacting, and when a specific mode is excited, we would know exactly where (i.e., to which modes and how fast) the energy is moving. Herein we develop such a formalism to simulate mode energy transfers in MD, in hopes of opening the path to understanding and rational engineering of various phenomena involving phonons, ranging from thermal transport[25,26] and superconductivity[27], to chemical reactions[16,18,28] and phase transitions [20].

**Phonon dynamics: mode forces and energy transfer.**

To write the equations of motion entirely in terms of mode-mode interactions, we seek to write an expression for the force on a mode due to interactions with other modes. A rigorous basis for deriving the force on degrees of freedom, such as individual atoms or modes, begins with the system Hamiltonian (total system energy). From the Hamiltonian $H$, one may then apply

Hamilton's equations $\frac{dq_n}{dt} = \frac{\partial H}{\partial p_n}$ and $\frac{dp_n}{dt} = -\frac{\partial H}{\partial q_n}$ to get the equations of motion, where $q_n$ is a generalized coordinate (e.g., it could be the displacement of an atom in some Cartesian direction, or the amplitude of a single normal mode) and $p_n$ is a generalized momentum (e.g., it could be the velocity of an atom in some Cartesian direction, or the rate of change of a mode amplitude) of a single degree of freedom $n$.[29] We therefore seek to obtain the Hamiltonian in terms of mode coordinates, from which we will use Hamilton's equations to get the equations of motion for the normal modes.

To obtain the mode Hamiltonian, we must first write the total Hamiltonian of a solid in terms of Cartesian coordinates of individual atoms:

$$H = \frac{1}{2}\sum_{i,\alpha} m_i \left(\dot{u}_i^\alpha\right)^2 + \frac{1}{2!}\sum_{\substack{ij \\ \alpha\beta}} \Phi_{ij}^{\alpha\beta} u_i^\alpha u_j^\beta + \frac{1}{3!}\sum_{\substack{ijk \\ \alpha\beta\gamma}} \Psi_{ijk}^{\alpha\beta\gamma} u_i^\alpha u_j^\beta u_k^\gamma + \cdots \quad (1)$$

where the first term on the right-hand side is the system kinetic energy in terms of atom masses $m_i$ for each atom $i$, and velocities $\dot{u}_i^\alpha = \frac{du_i^\alpha}{dt}$, where $u_i^\alpha$ is the displacement of atom $i$ in the $\alpha$ Cartesian direction. The rest of the terms on the right-hand side of Equation 1 are the system potential energy written as a Taylor expansion about equilibrium, as commonly done in the lattice dynamics literature[11,30]. The first term in the potential involves 2nd order force constants $\Phi_{ij}^{\alpha\beta}$ between atoms $i$ and $j$ in the $\alpha$ and $\beta$ Cartesian directions, the next term involves 3rd order force constants $\Psi_{ijk}^{\alpha\beta\gamma}$ between atoms $i, j, k$ in the $\alpha, \beta, \gamma$ Cartesian directions, and so forth. In this representation (Cartesian space), the generalized coordinates are $u_i^\alpha$ and generalized momenta are $m_i \dot{u}_i^\alpha$. Hamilton's equation $\frac{dp_n}{dt} = -\frac{\partial H}{\partial q_n}$ therefore gives the force on atom $i$ in the $\alpha$ Cartesian direction as

$$\frac{d(m_i \dot{u}_i^\alpha)}{dt} = -\frac{\partial H}{\partial u_i^\alpha} = F_i^\alpha = -\sum_{j,\beta} \Phi_{ij}^{\alpha\beta} u_j^\beta - \frac{1}{2} \sum_{jk,\beta\gamma} \Psi_{ijk}^{\alpha\beta\gamma} u_j^\beta u_k^\gamma - \cdots \quad (2)$$

which is the expression commonly used in the lattice dynamics literature to model the forces on atoms in solids[11]. The harmonic part of the equation of motion, $m_i \ddot{u}_i^\alpha = -\sum_{j,\beta} \Phi_{ij}^{\alpha\beta} u_j^\beta$, may be written in the form of an eigenvalue problem which defines the normal mode frequencies and eigenvectors of a solid; the form we use to obtain mode frequencies and eigenvectors is described in the Methods section.

We will follow the procedure of obtaining the force on individual atoms (Equation 2), in terms of interactions with other atoms, to obtain the force on individual modes, in terms of interactions with other modes. First, we transform that Hamiltonian from Cartesian coordinates to normal mode coordinates using the Fourier series representation[31,32] of atomic displacements $u_i^\alpha = \frac{1}{\sqrt{m_i}} \sum_n X_n e_{ni}^\alpha$ where $X_n$ is the mode amplitude (or generalized coordinate in the context of Hamiltonian mechanics) of mode $n$, and $e_{ni}^\alpha$ is the eigenvector component of atom $i$ in mode $n$ in the $\alpha$ Cartesian direction. Time-differentiating gives a similar expression for atomic velocities $\dot{u}_i^\alpha = \frac{1}{\sqrt{m_i}} \sum_n \dot{X}_n e_{ni}^\alpha$, where $\dot{X}_n$ is the mode velocity (or generalized momentum in the context of Hamiltonian mechanics) of mode $n$. Substituting these Fourier series representations of atomic displacements and velocities into the Hamiltonian of Equation 1, and rearranging with simple algebra, yields the Hamiltonian in terms of normal mode coordinates

$$H = \frac{1}{2}\sum_n \dot{X}_n^2 + \frac{1}{2!}\sum_n \omega_n^2 X_n^2 + \frac{1}{3!}\sum_{nml} K_{nml} X_n X_m X_l + \cdots \qquad (3)$$

which is the same form commonly used in the study of normal modes in the literature[33]. The indices $n, m, l$ represent different modes. The first term on the right-hand side is the system kinetic energy. The second term is the harmonic part of the potential energy and arises from our substitution of atomic displacements: $\sum_{\substack{ij \\ \alpha\beta}} \Phi_{ij}^{\alpha\beta} u_i^\alpha u_j^\beta = \sum_{nm} \sum_{\substack{ij \\ \alpha\beta}} e_{ni}^\alpha \frac{\Phi_{ij}^{\alpha\beta}}{\sqrt{m_i m_j}} e_{mj}^\beta X_n X_m = \sum_n \omega_n^2 X_n^2$ noting that the eigenvectors diagonalize the dynamical matrix[32], and $\omega_n$ is the frequency of mode $n$. Algebra shows that the 3$^{rd}$ order constants $K_{nml}$ are given by

$$K_{nml} = \sum_{ijk} \sum_{\alpha\beta\gamma} \frac{\Psi_{ijk}^{\alpha\beta\gamma} e_{ni}^\alpha e_{mj}^\beta e_{lk}^\gamma}{\sqrt{m_i m_j m_k}} \qquad (4)$$

which determine the strength of anharmonic interaction or coupling between modes; we therefore refer to these constants as the 3$^{rd}$ order mode coupling constants (MCC3s). Using this mode Hamiltonian, we may now apply Hamilton's equation $\frac{dp_n}{dt} = -\frac{\partial H}{\partial q_n}$ to get the force $F_n$ on mode $n$

$$\frac{d\dot{X}_n}{dt} = -\frac{\partial H}{\partial X_n} = F_n = -\omega_n^2 X_n - \frac{1}{2}\sum_{nml} K_{nml} X_m X_l - \cdots \qquad (5)$$

which is the expression used to study the dynamics of modes excited by lasers in the literature[34,35]. Like how the atom equation of motion in Equation 2 shows how interatomic interactions produce forces on atoms, the mode equation of motion shows how inter-mode interactions produce forces on modes. In the harmonic limit, however, the force on a mode is solely determined by its own amplitude $X_n$ times the frequency squared, since normal modes are defined so they do not interact in the harmonic limit of small vibrations.

The utility of knowing inter-mode forces via Equation 5 is that we may use them to calculate power transfer between modes. The usual expression for power transfer is given by force times velocity, and for modes this is derived by considering the time-rate of energy change for some mode $n$; this is $\frac{dH_n}{dt}$ where $H_n$ is the Hamiltonian of mode $n$, or a single term in the sums Equation 3. Since $H_n$ is a function of only the generalized coordinates and momenta, its total time-derivative is given by $\frac{dH_n}{dt} = \{H_n, H\} = \sum_m \left( \frac{\partial H_n}{\partial X_m} \frac{\partial H}{\partial \dot{X}_m} - \frac{\partial H_n}{\partial \dot{X}_m} \frac{\partial H}{\partial X_m} \right)$, which is the Poisson bracket of $H_n$ and the total system Hamiltonian $H$. This Poisson bracket is simplified by noting that $\frac{\partial H}{\partial \dot{X}_m} = \dot{X}_m$ and $\frac{\partial H_n}{\partial \dot{X}_m} = \delta_{nm} \dot{X}_m$, where $\delta_{nm}$ is the Kronecker delta. Using these substitutions, and the fact that $H = \sum_m H_m$, the total time-derivative may be written as $\frac{dH_n}{dt} = \sum_m \left( \frac{\partial H_n}{\partial X_m} \dot{X}_m - \frac{\partial H_m}{\partial X_n} \dot{X}_n \right)$. From the mode force of Equation 5, the remaining derivatives are the inter-mode forces given by $F_{nm} = -\frac{\partial H_n}{\partial X_m} = -\frac{1}{2} \sum_l K_{nml} X_m X_l$ for 3$^{\text{rd}}$ order anharmonic forces. The time-rate of change of a mode's energy is therefore given by $\frac{dH_n}{dt} = \sum_m \dot{Q}_{nm}$, where

$$\dot{Q}_{nm} = F_{nm} \dot{X}_n - F_{mn} \dot{X}_m \tag{6}$$

is the net rate of power transfer between modes $n$ and $m$. The power transfer between modes via Equation 6 is readily calculated in MD by simply storing the mode coupling constants, and using mode amplitudes and velocities at each timestep to evaluate the expression.

Before using Equation 6 to study mode energy transfer, we first verify that it correctly models energy transfer between modes, and that the calculated mode forces are correct, by considering a simple case. Our simple scenario involves exciting a localized interface mode, as shown in Figure 1, in an 800-atom silicon-germanium (Si-Ge) superlattice modeled by the Tersoff potential[36]. This excitation occurs while all other modes are at zero Kelvin, and we observe the resulting energy relaxation pathways. We found that after some time in the MD simulation, the energy of the initially excited interface mode (Mode 1) decreased, while two other modes (Mode 2 and Mode 3) increased in energy. This process is shown as a function of time in Figure 2.

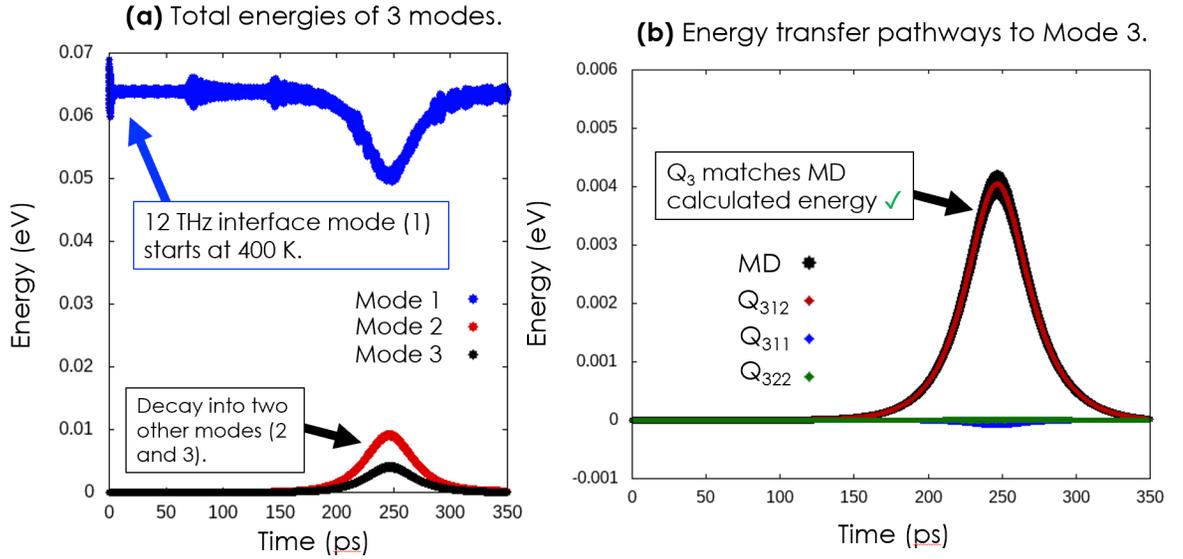

**Figure 2.** A MD simulation of the scenario in which a 12 THz interface mode is initialized with 400 K of energy, and all other modes in an 800-atom Si/Ge superlattice system are static with zero energy/amplitude. The dynamics were simulated with a Tersoff potential[36]. **(a)** During the MD simulation it is observed that the interface mode (denoted by Mode 1) decays in energy while two other modes (labeled Mode 2 and Mode 3) rise in energy. The mode energies are calculated via $E_n = \frac{1}{2}\omega_n^2 X_n^2 + \frac{1}{2}\dot{X}_n^2$. **(b)** We calculate the components of Mode 3's energy transfer by time-integrating Equation 6, and show its non-zero components ($Q_{312}$, $Q_{311}$, and $Q_{322}$) to show that the sum of these components equals the total energy calculated in MD. In fact, the energy transfer to Mode 3 is nearly entirely determined by a single 3-mode interaction $Q_{312}$ with Modes 1 and 2. This verifies that Equation 6 and the MCC3s are correct and self-consistent.

How do Modes 2 and 3 increase in energy, when they initially have zero energy/amplitude? If the system was purely harmonic, Modes 2 and 3 would never get excited because their forces $F_2 = -\omega_2^2 X_2$ and $F_3 = -\omega_3^2 X_3$ would remain zero since their initial amplitudes are zero, i.e., $X_n = 0$. Our system, however, is modeled with an anharmonic potential, so there are finite forces on Modes 2 and 3 of the form $F_2 = -K_{211} X_1 X_1$ and $F_3 = -K_{311} X_1 X_1$, due to the amplitude $X_1$ of Mode 1. These forces excite Modes 2 and 3 enough to activate a stronger energy transfer pathway $\dot{Q}_{312} = F_{312} \dot{X}_3$, where $F_{312} = -K_{312} X_1 X_2$, and this further excites Mode 3. Our goal here is to show that our method of calculating mode energy transfer is self-consistent, so we did not investigate why Mode 1 excites Modes 2 and 3 without exciting other modes. This approach could be used, however, to study phenomena such as the Fermi-Pasta-Ulam-Tsingou problem which saw similar behavior[37].

To illustrate that this approach is correct, we focus on Mode 3 which is one of the modes that gains energy, and we compare its calculated MD energy $E_3 = \frac{1}{2}\omega_3^2 X_3^2 + \frac{1}{2}\dot{X}_3^2$ to the sum of energy transfers due to other modes $Q_3 = Q_{311} + Q_{312} + Q_{313} + Q_{322} + Q_{323} + Q_{333}$, and find that this sum is equal to the total mode energy at each time step in the simulation. This confirms the validity of Equation 6, at every instant in time, and all the equations it depends on. In essence, Equations 4-6 mathematically answer the earlier question about the nature of mode-mode interactions and energy transfer; modes exert anharmonic forces on each other, just like atoms exert forces on each other, and the magnitude of these forces depends on other mode amplitudes and the degree of eigenvector overlap with MCC3s. For energy transfer, one may multiply this mode force with mode velocity to calculate power transferred to or from other modes, as shown in Eqauation 6. We call this method phonon or normal mode "dynamics" because it elucidates the dynamics of phonon/mode energy transfer processes in real-time.

**An example case of phonon-interface scattering.**

While we showed a simple scenario of mode energy transfer when all other modes are at 0 K to illustrate the concept of phonon dynamics, more important is the extension of this method to more realistic scenarios in which all other modes are at finite temperature, so that all anharmonic energy transfer pathways are activated. An example case we apply our method to in this paper is the collision of phonon wave-packets with an interface in a Si-Ge superlattice. For this example, we modeled the two materials with a neural network potential[38] trained on *ab initio* forces.

For this example, we focus on crystalline silicon (c-Si) phonon wave-packets with a frequency above the maximum frequency (9.3 THz) on the crystalline germanium (c-Ge) side, so that they cannot directly transmit through the interface, because no such frequencies exist in c-Ge. The idea is that since frequencies above 9.3 THz do not exist in c-Ge, the c-Si wave-packets must transfer energy through anharmonic pathways in order to move energy to the other side of the interface. This energy transport process is often termed "inelastic scattering" because the phonon must transfer energy to lower frequencies on the other side of the interface, which is a process that has been studied thoroughly in the literature[24,39,40]. Questions about the dynamics of inelastic phonon-interface interactions, however, remain unanswered. Notably, what mode energy transfers occur when a high-frequency phonon wave-packet collides with an interface? Are there preferred energy transfer pathways, or is energy distributed more evenly across the various interaction channels, since equipartition tends toward some uniformity in mode amplitudes? Others have suggested that pre-interface energy transfer dominates the wave-packet scattering[41], or that inelastic (anharmonic) transmission is significant[24,42,43], but the real-time dynamics of such an event remain unresolved. To answer these questions, we use the phonon dynamics method to study an inelastic phonon scattering event with unprecedented detail by observing, in real-time, what exactly happens when a phonon wave-packet collides with an interface at finite temperature. This real-time observation of mode energy transfer highlights the main advance, as it unveils the physics involved in mode energy transfer processes as they occur on femtosecond timescales accessible by MD simulations.

To answer the preceding questions, we focus on the anharmonic interaction pathways of longitudinal acoustic (LA) phonons launched from silicon to germanium in a Si-Ge superlattice, as a simple example. We focus on LA phonons since they are the most dominant energy carriers in c-Si due to their high group velocity[11]. To study phonon transmission in MD, we launched phonon wave-packets from the c-Si side, with frequencies higher than the c-Ge maximum frequency of 9.3 THz, to the c-Ge side. This is similar to the wave-packet method of Schelling et al.[2], but with one important change, needed to observe anharmonic effects, namely the rest of the system is given a finite temperature of 300 K. If we were to not add background vibrations at 300 K, the wave-packet could not transmit energy across the interface, because no such frequency exists on the germanium side. However, with phonon dynamics we can observe, in real-time, what happens when a phonon collides with an interface at finite temperature when all anharmonic transport pathways are activated, so that anharmonic scattering pathways may be observed. Hence, the approach taken herein can be termed a "finite-temperature wave-packet method". The details associated with generating the wave-packets in a finite-temperature environment are provided in the Methods section.

After initializing a phonon wave-packet in a finite temperature environment, we track its location and calculate its energy transfer pathways throughout time to determine what happens as it hits the interface. One question that arises in this scenario, where there are background 300 K vibrations, is how does one determine the location of the wave packet? In Figure 3 we illustrate that a wave-packet can be somewhat discernable when looking at both the atomic displacement field in real-space, and in the energy spectrum. As Figure 3 illustrates, we define the wave-packet location $z_{WP}$ with a center of energy, by analogy to an object's center of gravity, via

$$z_{WP} = \frac{1}{\sum_{n,i} E_{ni}} \sum_{n,i} z_i E_{ni} \qquad (7)$$

where the sums are conducted over all modes $n$ that initially comprise the wave-packet, and atoms $i$ in those modes, so that $E_{ni} = \frac{1}{2}\omega_n^2 X_n X_{ni} + \frac{1}{2}\dot{X}_n \dot{X}_{ni}$ is the energy contribution of atom $i$ to mode $n$. The per-atom contributions to mode amplitudes and velocities are given by $X_{ni} = \sqrt{m_i}\mathbf{e}_{ni} \cdot \mathbf{u}_i$ and $\dot{X}_{ni} = \sqrt{m_i}\mathbf{e}_{ni} \cdot \mathbf{v}_i$, respectively, where $\mathbf{e}_{ni}$ is the mode $n$ eigenvector on atom $i$, and $\mathbf{u}_i$ and $\mathbf{v}_i$ are the displacement and velocity vectors respectively.

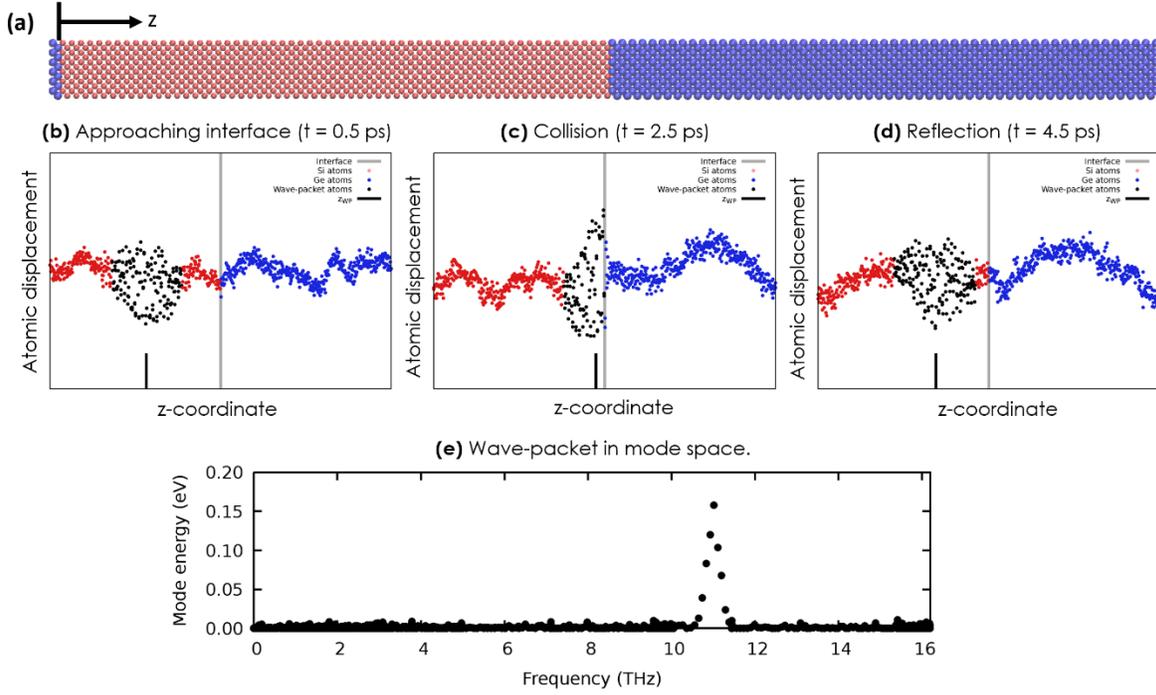

**Figure 3. (a)** The 4320-atom Si-Ge superlattice for which we perform wave-packet simulations. The wave-packets are launched from the Si side (red atoms) to the Ge side (blue atoms), in the direction of the defined z-coordinate. **(b)-(d)** illustration of an 11 THz LA wave-packet in a finite temperature environment, for illustration purposes. visualized as a real-space atomic displacement field at different times during a MD simulation; the black dots are displacements of all atoms in the system along the z-direction, the interface is shown by the blue lines and wave-packet position $z_{WP}$ is shown by the red line. The energy spectrum of all modes is shown in **(e)**, to show that the wave-packet has a notable peak-like signal. We seek to unveil the mechanisms of scattering in real-time; do significant energy transfers occur when the wave-packet hits the interface around 2.5 ps in (c)? Using our formalism, we can answer such questions.

To quantitatively study what happens when phonon wave-packets collide with an interface, we launched 10 LA wave-packets, from reduced wave-vector $k = 0.75$ to $k = 0.99$ in the [001] direction, from c-Si to the Si-Ge interface, while the total system temperature was 300 K. Since the MD simulation is conducted at finite temperature, one single trajectory cannot fully encapsulate what happens on average. Therefore, we performed 50 ensembles of each wave-packet simulation starting from different initial conditions (different random velocities which determine the background 300 K temperature), to make general conclusions about where energy is transferred on average. Here, it is important to distinguish this ensemble averaging from the time averaging that takes place in other approaches like GKMA[9] and ICMA[3]. In the approach we introduced herein, the time sequence of events, namely the collision of the wave packet with the interface is preserved across all ensembles. Thus, the approach introduced herein allows us to extract information and draw conclusions about not just what is happening, but when it happens, which is important for better understanding of energy transport processes. This is to mean, for example, that we can distinguish between which energy channels are activated, before, during and after the collision, since we now have access to the phonon dynamics.

We plotted the energy transfers to different categories of modes as a function of time in conjunction with the wave-packet center of energy (Equation 7), to quantitatively show what happens when the wave-packet hits the interface. Two of these wave-packet simulation results are shown in Figure 5. We note here that all wave-packets from $k = 0.75$ to $k = 0.99$ along the [001] direction of the dispersion curve exhibited similar behavior, so we only show the bounds $k = 0.75$ and $k = 0.99$ here.

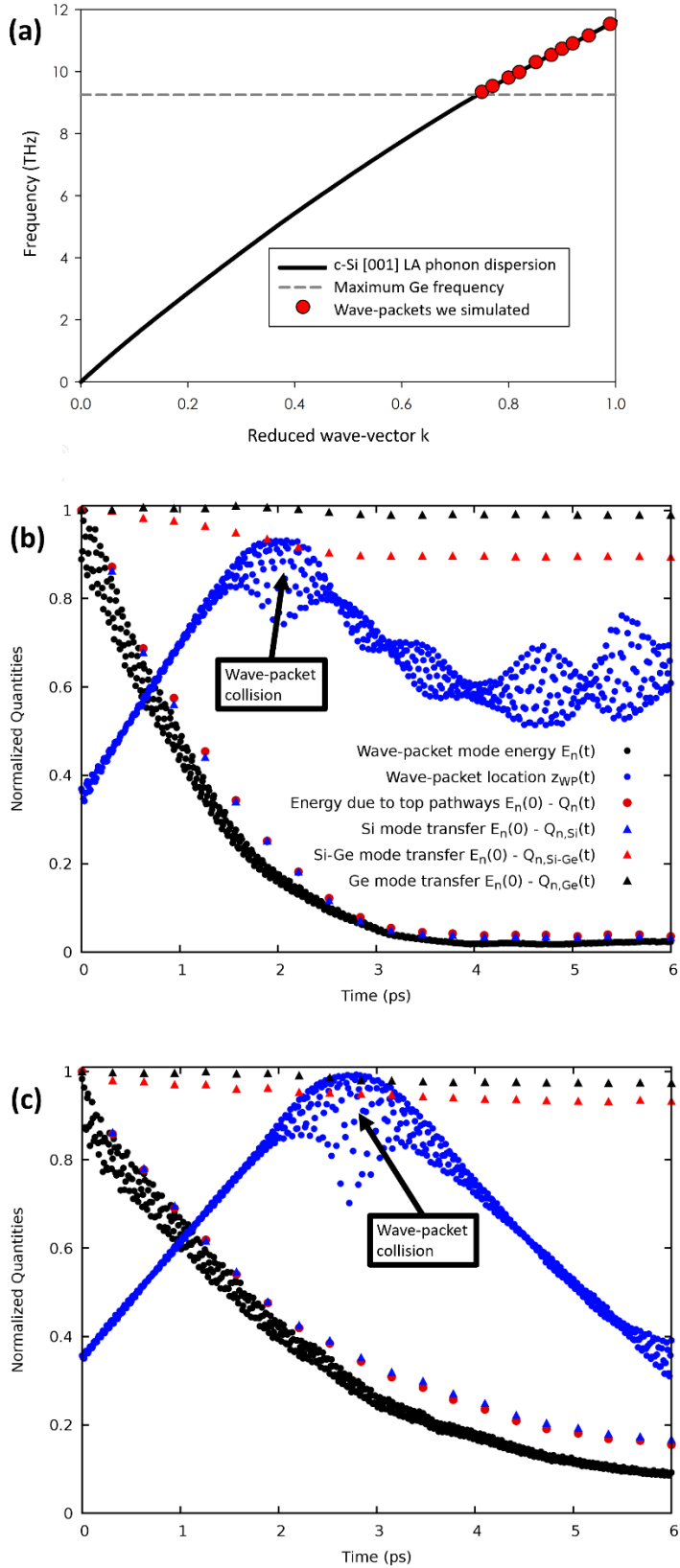

**Figure 4.** Finite temperature wave-packet simulation results. We note that each datapoint in **(b)** and **(c)** are average quantities over 50 ensembles starting from different initial conditions, so we are observing general behavior not particular to a unique set of initial conditions. The 10 wave-packets we simulated are shown by the red circles in **(a)** on the c-Si dispersion curve in the [001] direction, with the maximum germanium frequency as a horizontal line to show that these wave-packets cannot transmit into the c-Ge side of the Si/Ge interface. We show the results for two modes here: $k = 0.75$ and $k = 0.99$ in **(b)** and **(c)**, respectively. **(b)** and **(c)** show normalized quantities associated with the wave-packet as a function of time. Each dot/triangle represents an average over 50 ensembles, and the average of these quantities are plotted as a function of time to generalize about the behavior of phonon-interface scattering. The meaning of the legend is described in the following text.

In Figure 4, the black dots are the normalized ensemble average of the energy $E_n(t) = \frac{1}{2}\omega_n^2 X_n^2(t) + \frac{1}{2}\dot{X}_n^2(t)$ of the mode at the center wave-packet (i.e., in mode/frequency space), which typically has ~30-50% of the total wave-packet energy at any given time. The red dots are an ensemble average of the mode energy calculated from the energy transfers to other modes, $E_n(t) = E_n(0) - Q_n(t)$ according to energy transfers $Q_n = \sum_m Q_{nm}$. These are calculated by time-integrating Equation 6, but here we have only included the top 0.5% of pathways with the largest MCC3 magnitudes. Since the difference between the black dots (actual mode energy) and red dots (mode energy calculated with energy transfers restricted to the top 0.5% strongest pathways) is within 10% for all 10 wave-packets, we conclude that the majority of energy transfer occurs through the top 0.5% of pathways with the largest MCC3 values. The blue dots in Figure 4 are the center of energy defined in Equation 7, normalized such that a value near unity means that the wave-packet is near the interface; this is why $z_{WP}$ increases until it achieves a value near unity, where it collides with the interface, and then decreases due to reflection away from the interface in the opposite direction. The triangles represent energy transfers to different types of modes, defined again as $E_n(0) - Q_n(t)$ where $Q_n = \sum_{ml} Q_{nml}$ but the sum over $ml$ includes only modes in a specific group: $Q_{n,Si}$ where $ml$ runs over modes which have eigenvectors primarily on the Si side only, $Q_{n,Si-Ge}$ where $ml$ runs over modes which have eigenvectors primarily in both the Si and Ge (e.g., interface modes or extended modes[44]), and $Q_{n,Ge}$ where $ml$ runs over modes which have eigenvectors primarily on the Ge side only.

Figure 4 answers questions we raised earlier, namely what happens when a wave-packet collides with an interface at finite temperature? Nothing in particular happens in terms of energy transmission across the interface; on average 1-2% of the energy transmits directly via interactions involving Ge modes, while most of the energy is exchanged with Si modes before and after the interface collision occurs. About 10% of the energy, however, is transferred to modes with eigenvectors that exist in both c-Si and c-Ge (red triangles in Figure 4). This includes localized interface modes as shown in Figure 1, or delocalized fully extended modes, although there are more of the latter than the former (we found ~30 localized interface modes, but hundreds of fully extended modes); this suggests that delocalized fully extended modes are therefore a significant pathway/bridge for direct inelastic energy transfer from c-Si to c-Ge.

We also raised the questions: Are there preferred energy transfer pathways, and if so, why? The answer to this question lies in the fact that the sum of the calculated energy transfers via time-integrating Equation 6, shown at each time by the red dots in Figure 4, consistently agreed (within 10%) with the total mode energy calculated in MD for all 10 wave-packet simulations. This means that about 90% of the mode energy transfers come from the top 0.5% of interactions associated with the largest MCC3 magnitudes, which then suggests that MCC3s can serve as a descriptor for total energy transfer. Here we note that there are a total of ~168 million different $3^{rd}$ order/3-phonon interaction terms for each mode in this system, yet only ~800,000 of them (or xx%) are responsible for almost all of the heat transfer. This is surprising, given the wide range of possible mode-mode couplings. With systems operating not far from equilibrium, equipartition would cause the mode amplitudes to be rather evenly excited, especially in a classical MD simulation. However, what seems to matter most is the wide range of values that the MCC3s take on, which span X orders of magnitude. In this sense, the matrix of MCC3s is sparse, just like a dynamical matrix that details the interatomic interactions. Another related question is whether the coupling constants also serve as a descriptor for where the energy goes. Do the largest MCC3s by magnitude correspond to the largest energy transfer pathways? To answer these questions, we may compare the magnitude of MCC3s with the actual energy transfer pathways during the wave-packet simulation.

**Visualizing energy transfer pathways – where does the energy go?**

While Figure 4 shows which categories (Si, Si-Ge, or Ge) of modes the energy is exchanged with, we seek to determine exactly where the energy is distributed among all possible pathways. The energy transfer matrix $Q_{nml}$ for each mode $n$ (i.e., which runs over the two remaining indices $m$ and $l$) was calculated by time-integrating the power transfer according to Equation 6 from 0 ps to 6 ps. This interval corresponds to the time that the wave-packets have attenuated completely away from the interface in Figure 4. A comparison of the energy transfer matrix $Q_{nml}$ to the MCC3 matrix $K_{nml}$ for the same two wave-packet modes ($k=0.75$ and $k=0.99$) are shown in Figure 5 as a 2D map representing the magnitude of energy transfers $Q_{nml}$ and MCC3s $K_{nml}$. We found similar results for all 10 wave-packet modes that we simulated from $k=0.75$ to $k=0.99$.

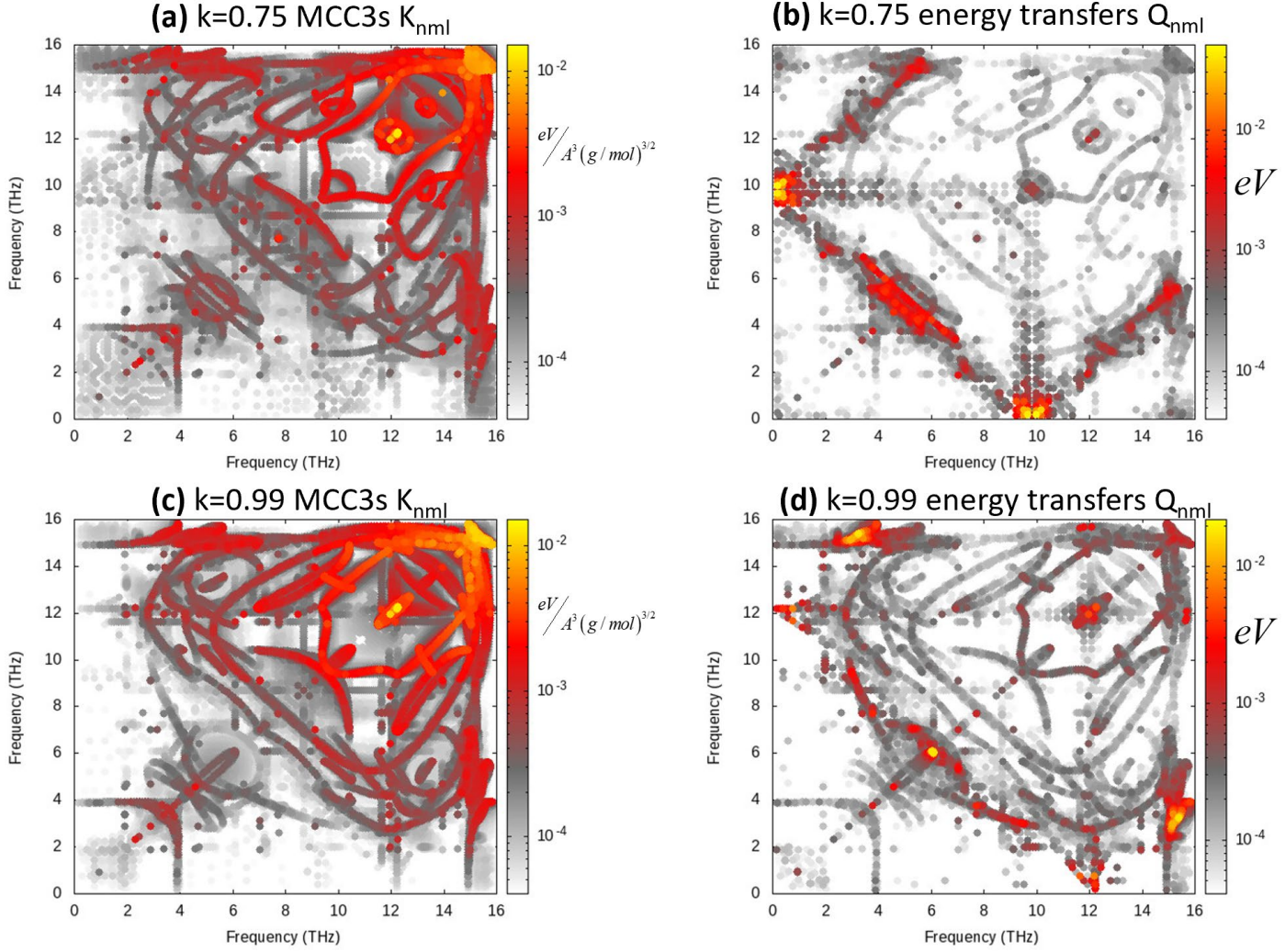

**Figure 5.** Comparison of MCC3s and energy transfers between all modes, for the two wave-packet modes at $k=0.75$ and $k=0.99$ on the dispersion curve in Figure 4. The energy transfer plots **(b)** and **(d)** are calculated by time-integrating Equation 6, where the final time is the end of the 6 ps wave-packet simulation illustrated in Figure 4. **(a)** Shows the MCC3s $K_{nml}$ given by Equation 4, where $n$ is the wave-packet mode at $k = 0.75$ and we plot the MCC3 magnitude $K_{nml}$ where $m$ and $l$ are represented as frequencies. **(b)** Shows the energy transfers $Q_{nml}$ given by time-integrating Equation 6, where $n$ is the wave-packet mode at $k = 0.75$ and we plot the energy transfer magnitude $Q_{nml}$ where $m$ and $l$ are represented as frequencies. Similarly, **(c)**-**(d)** show the same comparison for the $k = 0.99$ wave-packet mode.

Figure 5 shows that the energy transfers $Q_{nml}$ to all frequencies show some similar qualitative features, i.e., the shapes/non-white spaces are similar. However, the magnitudes exhibit different features compared to the MCC3s $K_{nml}$. Notably, the largest energy transfers do not necessarily correspond to the largest MCC3 values, as the preferred pathways only become apparent by simulating the phonon dynamics. These results answer another question that was raised: Are certain mode energy transfer pathways preferred? The energy transfer maps of Figure 5 suggest so, as it should be noted that the plots are shown on log scale. This indicates that some channels are much more strongly favored than others. While the largest MCC3s may not correspond exactly to the most preferred energy transfer pathways, all the dynamically preferred pathways are still within the top 0.5% of MCC3s. This suggests that the MCC3s are an important descriptor that can be used to filter out the overwhelming majority of channels, which are not active, as the MCC3s for this system span ~20 orders or magnitude. This large degree of variability in the MCC3 magnitudes renders the subgroup that has large magnitudes (i.e., the top 0.5%, which span the top 3-4 orders of magnitude) easy to obtain for relevant systems sizes of ~10,000-100,000 atoms. Despite the immense spectrum of possible phonon-phonon interactions, a matrix quantifying their interactions is sparse, just like interatomic interactions in a dynamical matrix. This finding has profound implications, as the MCC3s can serve as important descriptors for understanding a wide variety of phenomena in terms of different phonon coupling channels.

**Conclusion and future work.**

In this study we introduced a method for studying real-time phonon or normal mode dynamics. The method was derived rigorously from the Hamiltonian of a solid, showing that one can think of mode-mode interactons in a similar manner as atom-atom interactions. We proceeded to write the equations of motion for a normal mode, and identified the forces it experiences from other modes, which were derived by projection of the interatomic interactions onto the normal mode shapes. This then enabled simulation of the dynamics entirely in mode space, rather than real space, which led to a subsequent derivation of the energy transferred between modes that was shown to be fully self-consistent. The primary advance with this approach over other MD approaches is the ability to observe real-time vibrational energy transfer processes. We exhibited this real-time dynamics of modes by showing the energy transfer processes occuring when phonons collide with an interface at finite temperature, supporting the hypothesis that pre-interface scattering is a primary scattering mechanism in systems with interfaces[41]. In the process of studying this anharmonic phonon-interface scattering, we extended the traditional harmonic wave-packet method[2] to finite temperatures so that anharmonic scattering pathways may be observed.

We also found that the MCC3s span ~20 orders of magnitude, and the top ~0.5% of MCC3s, which comprise the top 3-4 orders of magnitude, are primarily responsible for transport. The MCC3s are therefore effectively sparse, just like interactions between atoms, which generally decay with distance; in the case of phonons, the interaction strength decays with the degree of spatial overlap in the eigenvectors. Thus, we identified the MCC3s as useful descriptor for understanding which mode interactions dominate, and point to a possible approach for engineering, i.e., by changing/nanostructuring a system to have large MCC3 values for increasing thermal conductance via inelastic transmission[26,45]. It is also important to note that the process of extracting modes, calculating their coupling constants, and calculating their forces and energy transfers in MD simulations is non-trivial. Therefore, to facilitate the use of the method presented herein, we have developed ModeCode[46], as an open-source massively parallel software package for such calculations. The generality of our method invites application to a variety of phenomena such as vibrational energy transfer mechanisms leading to chemical reactions[16], mass diffusion[22,47], protein conformation changes [48], catalysis[5-7], phase transitions[14] and other phenomena which are influenced by atom vibrations such as superconductivity[49].

**Methods.**

The frequencies and eigenvectors of normal modes for a solid in a supercell are defined by the harmonic part of the equation of motion, $m_i \ddot{u}_i^\alpha = -\sum_{j,\beta} \Phi_{ij}^{\alpha\beta} u_j^\beta$ from Equation 2. The exact solution of this 2$^{nd}$ order differential equation is a well-known in lattice dynamics[31,32] and takes the form of an eigenvalue problem. Usually this eigenvalue problem is solved for all wave-vectors in a periodic crystal to obtain phonon dispersion curves[11], but here we are interested in modes arising in non-crystalline systems, so we solve the eigenvalue problem at the gamma point (zero wave-vector) in a large supercell of atoms. After substituting general solutions $u_i^\alpha(t) = \frac{1}{\sqrt{m_i}} \sum_n X_n e_{ni}^\alpha e^{-i\omega_n t}$, where $e^{-i\omega_n t}$ is a time exponential showing the oscillatory behavior of a single mode, the harmonic equation of motion $m_i \ddot{u}_i^\alpha = -\sum_{j,\beta} \Phi_{ij}^{\alpha\beta} u_j^\beta$ can be written in the form of the eigenvalue equation[32]

$$\mathbf{e} \cdot \mathbf{\Omega} = \mathbf{D} \cdot \mathbf{e} \qquad (8)$$

where $\mathbf{e}$ are matrices of eigenvectors, with each column representing the eigenvector of a single mode, organized like[32]

$$\begin{matrix} e_{11}^x & e_{21}^x & e_{31}^x & \cdots & e_{3N,1}^x \\ e_{11}^y & e_{21}^y & e_{31}^y & \cdots & e_{3N,1}^y \\ e_{11}^z & e_{21}^z & e_{31}^z & \cdots & e_{3N,1}^z \\ e_{12}^x & e_{22}^x & e_{32}^x & \cdots & e_{3N,2}^x \\ e_{12}^y & e_{22}^y & e_{32}^y & \cdots & e_{3N,2}^y \\ \vdots & \vdots & \vdots & \ddots & \vdots \\ e_{1N}^z & e_{2N}^z & e_{2N}^z & \cdots & e_{3N,N}^y \end{matrix} \qquad (9)$$

where the eigenvector components $e_{ni}^\alpha$ are indexed by mode $n$, atom $i$, Cartesian direction $\alpha$, and the mode indices run from 1 to $3N$ ($N$ being the number of atoms). The $\boldsymbol{\Omega}$ in Equation 8 is a diagonal matrix of eigenvalues which are the squared frequencies of modes given by

$$\begin{pmatrix} \omega_1^2 & 0 & 0 & 0 \\ 0 & \omega_2^2 & 0 & 0 \\ 0 & 0 & \ddots & 0 \\ 0 & 0 & 0 & \omega_{3N}^2 \end{pmatrix} \qquad (10)$$

where $\omega_n$ is the frequency of mode $n$. The dynamical matrix $\mathbf{D}$ is an $N \times N$ array of smaller $3 \times 3$ matrices; the elements of these $3 \times 3$ blocks are given by

$$D_{ij}^{\alpha\beta} = \frac{\Phi_{ij}^{\alpha\beta}}{\sqrt{m_i m_j}} \qquad (11)$$

where the position of this element in the full dynamical matrix $\mathbf{D}$ is given by row $3(i-1)+\alpha$ and column $3(j-1)+\beta$.[32] We made the open-source program ModeCode[46] to aid in calculating this dynamical matrix and solving the eigenvalue problem of Equation 8 to get the mode frequencies and eigenvectors for large supercells of atoms. In our open-source code, we also provide LAMMPS files to aid in the calculation of mode amplitudes and velocities at each timestep in a MD simulation. We note here that obtaining the mode amplitudes and velocities at every timestep in a MD simulation brings an additional cost which involves a double loop over all atoms: one loop over all modes (3 times the number of atoms), and another loop over all atoms and Cartesian directions, to calculate the mode amplitudes and velocities $X_n = \sum_{i,\alpha} \sqrt{m_i} u_i^\alpha e_{ni}^\alpha$ and $\dot{X}_n = \sum_{i,\alpha} \sqrt{m_i} \dot{u}_i^\alpha e_{ni}^\alpha$, respectively.

The finite-temperature wave-packet method involves initializing a phonon wave-packet in a finite-temperature environment, so that many anharmonic energy transfer pathways are activated. To achieve this, we initialize modes associated with LA phonon wave-packets using the atomic displacements[50]

$$u_i^\alpha = A e_i^\alpha(k_0) \exp\left[i k_0 (z_i - z_0)\right] \exp\left[-\frac{(z_i - z_0)^2}{\xi^2}\right] \qquad (12)$$

where $A$ is an amplitude representing the magnitude of displacement, $e_i^\alpha(k_0)$ is the polarization vector of mode eigenvector components, $k_0$ is the reduced wave-vector, $z_i$ is the Cartesian $z$ coordinate of atom $i$ with $z_0$ as the initial wave-packet position in real-space, and $\xi$ is the spatial extent of the wave-packet. It is important to note that Equation 8 along with its time derivative (atomic velocity $v_i^\alpha$) constitute a wave-packet which gives a unique signature of mode amplitude and velocity through the Fourier series representations $X_n = \sum_{i,\alpha} \sqrt{m_i} u_i^\alpha e_{ni}^\alpha$ and $\dot{X}_n = \sum_{i,\alpha} \sqrt{m_i} v_i^\alpha e_{ni}^\alpha$. Using these unique combinations of mode amplitude and velocity which constitute a wave-packet, we force a wave-packet to exist in a finite temperature environment by first initializing the system with randomized velocities, and then recalculating positions and velocities so that $u_i^\alpha = \frac{1}{\sqrt{m_i}} \sum_n X_n e_{ni}^\alpha$ and $v_i^\alpha = \frac{1}{\sqrt{m_i}} \sum_n \dot{X}_n e_{ni}^\alpha$, where $X_n$ and $\dot{X}_n$ are the mode amplitudes and velocities associated with the wave-packet. For all wave-packet simulations, we use a 4320-atom Si-Ge superlattice modelled by a neural network potential[51] that was trained against *ab initio* forces and performed MD simulations using LAMMPS[52].

**Competing Interests**

The authors declare no competing financial or non-financial interests.

**Acknowledgements**

The authors would like to thank ONR MURI (N00014-18-1-2429) for the financial support. AR was supported by the National Science Foundation Graduate Research Fellowship under Grant No. 1122374. Any opinion, findings, and conclusions or recommendations expressed in this material are those of the authors and do not necessarily reflect the views of the National Science Foundation.

**Author Contributions**

AR established the formalism for describing energy transfer between modes and created the Mode Code program for generally extracting modes and calculating their coupling constants. AH served as an advisor on this project and helped developed the phonon dynamics formalism.

**Data Availability**

The raw data required to reproduce these findings are available to download from https://github.com/rohskopf/ModeCode. The processed data required to reproduce these findings are available to download from https://github.com/rohskopf/ModeCode.